\documentclass[11pt]{article}

\usepackage[margin=1in]{geometry}
\usepackage[utf8]{inputenc}
\usepackage[T1]{fontenc}
\usepackage{hyperref}
\usepackage{url}
\usepackage{booktabs}
\usepackage[numbers]{natbib}
\usepackage{amsfonts}
\usepackage{amsmath}
\usepackage{amssymb}
\usepackage{amsthm}
\usepackage{mathtools}
\usepackage{graphicx}
\usepackage{microtype}
\usepackage{xcolor}
\usepackage{siunitx}

\graphicspath{{figures/}}
\hypersetup{colorlinks=true,linkcolor=blue,citecolor=blue,urlcolor=blue}
\newtheorem{lemma}{Lemma}

\title{ConRad: Efficient Conformal Prediction for Radiomics}

\author{%
  Matt Y. Cheung, Ashok Veeraraghavan, Guha Balakrishnan \\
  Department of Electrical \& Computer Engineering, Rice University \\
}

\begin{document}

\maketitle

\begin{abstract}
Radiomic features derived from medical images and segmentation masks are used to support decision making in clinical imaging pipelines.
In practice, these features are often computed from predicted masks, but segmentation models can be overconfident or poorly calibrated, making derived measurements appear more reliable than they are.
Conformal prediction (CP) provides distribution-free prediction intervals with finite-sample marginal coverage guarantees, but black-box intervals for segmentation-derived radiomics can be inefficient because they ignore test-time information about image appearance, mask geometry, and segmentation uncertainty. We propose \emph{ConRad}, a conformal framework for scalar radiomic targets that uses covariates derived from the predicted mask, input image, predicted radiomics, and boundary uncertainty to construct adaptive intervals while maintaining coverage. 
Across five 2D medical imaging datasets and 171 retained radiomic targets, we show that ConRad improves feature-level efficiency compared to baselines while maintaining near-nominal empirical coverage. Ablation results further indicate that segmentation boundary uncertainty features are the largest contributors to interval efficiency. 
\end{abstract}

\section{Introduction}

Radiomic measurements derived from segmentation masks are increasingly used as downstream quantities in medical imaging pipelines. 
These measurements, including shape, intensity, and texture features, are often computed from masks predicted by automated segmentation models and used for downstream analysis and decision making.
Thus, uncertainty quantification (UQ) is important in high stakes scenarios. A natural approach to uncertainty quantification (UQ) is to treat each radiomic measurement as a scalar prediction target and construct an interval directly around the predicted radiomic value. 
While this black-box view is reasonable, it ignores how the measurement is produced: the radiomic value is not an arbitrary scalar output, but a deterministic function of the input image and the predicted segmentation mask. As a result, direct UQ methods may leave efficiency gains on the table by failing to use information contained in the predicted mask, such as object size, topology, boundary location, and segmentation uncertainty. 
We therefore ask whether test-time information from the segmentation output can be used to construct narrower uncertainty intervals for segmentation-derived radiomics while preserving formal coverage guarantees.

Conformal prediction (CP) provides a statistically principled UQ framework with finite-sample guarantees under exchangeability~\cite{vovk2005algorithmic,angelopoulos2021gentle,shafer2008tutorial,fontana2023conformal}. Split CP uses a held-out calibration set to construct prediction sets for new test examples and guarantees marginal coverage at level $1-\alpha$. Existing CP methods for segmentation typically focus on pixel- or voxel-level uncertainty sets~\cite{mossina2025conformal,brunekreef2024kandinsky,viti2025consign,elyassirad2025conseg}. 
Although such sets can characterize contour uncertainty, spatial uncertainty maps do not directly provide calibrated intervals for downstream radiomic measurements. Recent metric-based CP methods~\cite{cheung2025compass,cheung2026efficient,cheung2026bias} construct conformal intervals for segmentation-derived metrics by leveraging pipeline-specific structure, such as U-Net features for segmentation~\cite{cheung2025compass} or deformation fields for registration~\cite{cheung2026efficient}. 
These methods show that segmentation-specific information can improve interval efficiency, but they have mainly focused on area- or volume-like quantities. 
It remains unclear how to obtain efficient conformal intervals for non-volumetric radiomic targets, whose values may depend on shape, intensity distribution, texture, and boundary uncertainty.

\begin{figure}[t!]
    \centering
    \includegraphics[width=\linewidth]{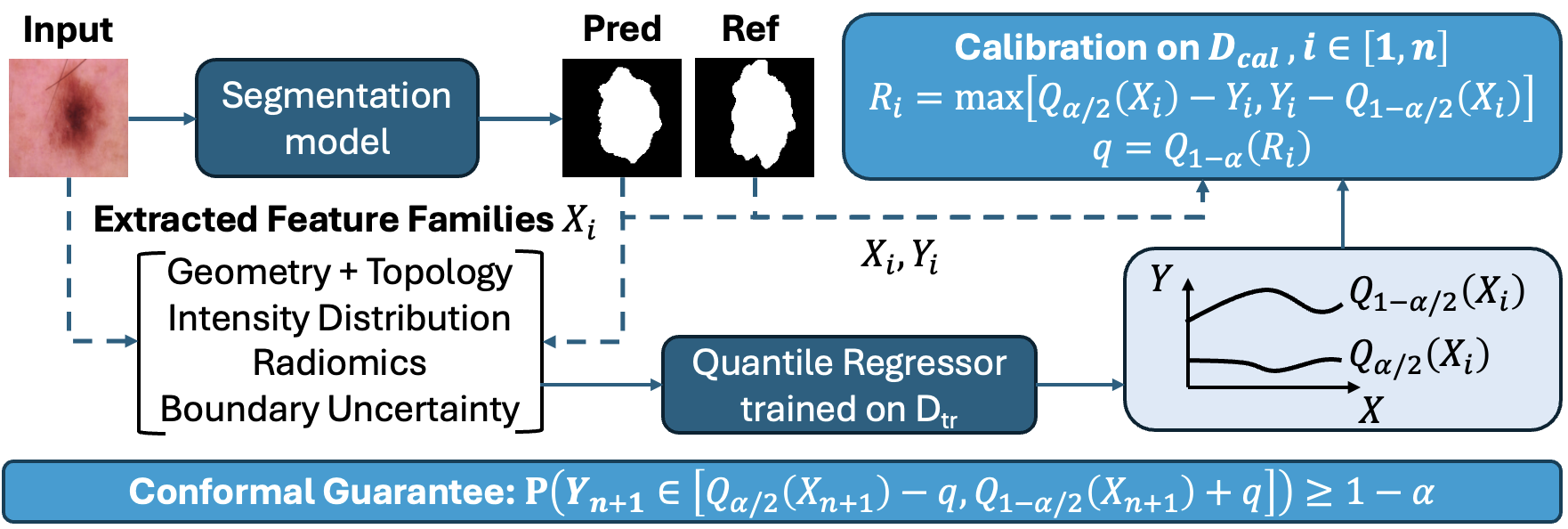}
    \caption{\textbf{Overview of ConRad.} For each radiomic target, a training dataset $D_{\mathrm{tr}}$ is used to fit lower and upper conditional quantile models $\hat Q_{\alpha/2}(X_i)$ and $\hat Q_{1-\alpha/2}(X_i)$ for the reference target $Y_i$. The features $X_i$ are available at test time and are derived from the predicted radiomic value, predicted mask, image summaries, and predicted probability map. A disjoint calibration dataset $D_{\mathrm{cal}}$ is then used to compute CQR nonconformity scores and adjust the interval by the conformal quantile $\hat q$. Under exchangeability, the resulting interval for a new case has marginal coverage at least $1-\alpha$.}
    \label{fig:overview}
\end{figure}

We propose \emph{ConRad}, an efficient CP framework for scalar radiomic targets derived from predicted segmentation masks. 
ConRad uses conformalized quantile regression (CQR) to learn adaptive intervals from test-time covariates derived from the predicted mask, input image, predicted radiomic values, and boundary uncertainty. This allows interval width to vary with case difficulty while retaining split-conformal coverage after calibration. In contrast to direct black-box intervals around predicted radiomic values, ConRad explicitly uses information about the segmentation output and its uncertainty to improve interval efficiency.

Because radiomic features differ substantially in scale, stability, and interpretation, raw interval widths cannot be averaged meaningfully across targets. 
We therefore evaluate efficiency using target-level win rates and symmetric relative width improvement against the strongest baseline, rather than pooled raw interval width. 
Across five 2D medical imaging datasets and 171 retained radiomic targets, ConRad improves feature-level interval efficiency compared with split conformal prediction and CQR baselines while maintaining near-nominal empirical coverage. 
A grouped ablation analysis shows that boundary uncertainty is the largest contributor to interval-efficiency gains, suggesting that segmentation uncertainty near object boundaries provides complementary information beyond generic image and mask summaries.

Our contributions are twofold. First, we formulate ConRad, a feature-adaptive conformal framework for scalar radiomic targets derived from predicted segmentation masks. 
Second, we introduce a scale-normalized, target-level evaluation protocol for comparing interval efficiency across heterogeneous radiomic targets. 
Together, these contributions extend metric-level conformal uncertainty quantification beyond volume-based segmentation measurements to broader radiomic measurements computed from medical image segmentation masks.

\section{Conformal Prediction for Radiomic Targets}
\label{sec:method}

We first define the radiomics prediction problem and the conformalized quantile regression (CQR) procedure used to construct prediction intervals. We then describe the filtering procedure used to retain stable radiomic targets and the four families of test-time features used by ConRad: geometry and topology, intensity distribution, predicted radiomics, and boundary uncertainty. Figure~\ref{fig:overview} summarizes the training, calibration, and test-time workflow.

\subsection{Setup}

Let $A_i$ denote the image for case $i$, let $\hat M_i$ be the predicted segmentation available at inference time, and let $M_i$ be the reference segmentation used only for training, calibration, and evaluation. We use a radiomics extractor $\rho:\mathcal A\times\mathcal M\rightarrow\mathbb{R}$ to map an image-mask pair to a scalar radiomic target. For a retained target, the predicted and reference radiomic values are
$\hat{Y}_i = \rho(A_i, \hat M_i) \in \mathbb{R}$ and $Y_i = \rho(A_i, M_i) \in \mathbb{R}$.
In addition to $\hat Y_i$, we observe auxiliary features $\hat Z_i$ derived only from test-time information: the image $A_i$, predicted mask $\hat M_i$, and predicted foreground probability map. We collect all features in $X_i=(\hat Y_i,\hat Z_i)$ where $\hat Z_i$ includes geometry/topology summaries, intensity-distribution summaries, predicted-radiomics summaries, and boundary-uncertainty summaries. For each radiomic target, our goal is to construct an interval-valued predictor $S(X_i)$ that is narrow while satisfying finite-sample marginal coverage, $\Pr\{Y_{\mathrm{test}}\in S(X_{\mathrm{test}})\}\ge 1-\alpha$ under exchangeability.

Given a calibration set $D_{\mathrm{cal}}=\{(X_i,Y_i)\}_{i=1}^n$ disjoint from the training set, and assuming calibration and test examples are exchangeable, CQR trains conditional lower and upper quantile models $\hat Q_{\alpha/2}(X)$ and $\hat Q_{1-\alpha/2}(X)$ for the reference radiomic target $Y$ on the training set. The trained quantile models are fixed before calibration. On the calibration set, we compute nonconformity scores $R_i=\max\{\hat Q_{\alpha/2}(X_i)-Y_i,\;Y_i-\hat Q_{1-\alpha/2}(X_i)\}$.
Let $\hat q$ be the $\lceil(1-\alpha)(n+1)\rceil$-th smallest value of $\{R_i\}_{i=1}^n$. The conformalized interval for a test case is
$S(X_{\mathrm{test}})=
\left[
\hat Q_{\alpha/2}(X_{\mathrm{test}})-\hat q,\;
\hat Q_{1-\alpha/2}(X_{\mathrm{test}})+\hat q
\right]$.
Thus, $\hat q$ is the calibration-based adjustment needed to achieve the target marginal coverage level.

\begin{lemma}
Let $\{(X_i,Y_i)\}_{i=1}^n$ and $(X_{\mathrm{test}},Y_{\mathrm{test}})$ be exchangeable after the quantile models have been trained on a disjoint training set. Define the CQR score for any candidate value $y$ as $R(X,y)=\max\{\hat Q_{\alpha/2}(X)-y,\; y-\hat Q_{1-\alpha/2}(X)\}$.
If $\hat q$ is the $\lceil(1-\alpha)(n+1)\rceil$-th order statistic of the calibration scores, then the conformal set $S(X_{\mathrm{test}})=\{y:R(X_{\mathrm{test}},y)\le \hat q\}$
satisfies $\Pr\{Y_{\mathrm{test}}\in S(X_{\mathrm{test}})\}\geq 1-\alpha$.
\end{lemma}
This is a standard split-conformal guarantee~\cite{vovk2005algorithmic,shafer2008tutorial,angelopoulos2021gentle}. It is marginal over exchangeable test examples and does not imply conditional coverage for every individual case or subgroup. The remainder of this section describes the retained radiomic targets and the test-time features used to adapt interval width.

\subsection{Target and feature Selection}

ConRad is designed for segmentation-derived radiomics, where prediction difficulty is target dependent rather than represented by a single scalar. A predicted mask may be reliable for one radiomic target and unreliable for another. 
We therefore use a wide variety of test-time features that can explain case difficulty for different radiomic targets. 
Separately, some radiomic targets become degenerate in small or homogeneous regions, producing missing values, near-constant values, or numerically unstable measurements. 
Including such targets would make coverage and interval-width comparisons reflect artifacts of feature extraction rather than meaningful uncertainty. 
Thus, a dataset-specific filtering step is necessary to keep the benchmark focused on radiomic quantities that are interpretable, measurable across enough cases, empirically variable, and suitable for fair conformal interval evaluation.

\textbf{Filtering Procedure.}
For each dataset, we selected radiomic targets using the same filtering rules applied separately within each dataset.
We first restricted attention to three target families: shape features, first-order intensity features, and texture or heterogeneity features. 
Shape features summarize lesion geometry and boundary morphology. 
First-order features summarize the intensity distribution inside the predicted region, and texture features summarize spatial heterogeneity. 
Targets with frequent non-finite values, too few usable observations, few distinct rounded reference values, or effectively zero variance in either predicted or reference form were removed. 
These filters ensured that interval coverage and width comparisons were not dominated by missing values, degenerate distributions, or numerical artifacts.
After target filtering, ConRad constructs test-time features derived from quantities available at inference time, including geometry and topology, intensity distribution, predicted radiomics, and boundary uncertainty features. 
We summarize the feature families below.

\textbf{Geometry and Topology.}
We summarize the basic geometry and topology of the predicted segmentation mask. 
We used the number of pixels in the predicted region, the fraction of the image occupied by the predicted mask, and the number of connected components in the predicted mask. 
These features indicate whether the predicted region is small or large, compact or fragmented, and therefore whether downstream radiomic measurements may be unstable.

\textbf{Intensity Distribution.}
We summarize the intensity distribution inside the predicted region of interest from the input image.
We used standard ROI intensity statistics: mean, standard deviation, minimum, maximum, and the 10th, 25th, 50th, 75th, and 90th percentiles.
These features describe the brightness, contrast, spread, and coarse intensity distribution of the segmented region.

\textbf{Predicted Radiomics.}
We included predicted radiomic features and retained radiomic targets from three radiomic feature families: 8 shape features (elongation, major axis length, maximum diameter, mesh surface, minor axis length, perimeter, perimeter-surface ratio, and pixel surface), 13 first-order intensity targets (10th percentile, 90th percentile, entropy, interquartile range, kurtosis, maximum, mean, median, minimum, range, skewness, uniformity, and variance), and 19 class-specific heterogeneity/texture targets from GLCM, GLDM, GLRLM, GLSZM, and NGTDM. 
We used PyRadiomics~\cite{van2017computational} to compute these.

\textbf{Boundary Uncertainty.}
We used probability statistics and entropy inside the predicted mask, within a mask boundary band, and globally, including mean, standard deviation, the 10th and 90th percentiles, and the fraction of pixels with foreground probability in $[0.4,0.6]$. 
We created the boundary band by taking the predicted binary mask and taking the difference between dilating and eroding the region by 1 pixel. 
These features capture whether the segmentation model is confident, uncertain near the boundary, or broadly uncertain across the image.
\section{Experiments}
\label{sec:experiments}

\textbf{Datasets and Targets.}
We evaluated ConRad on five 2D mask-annotated segmentation datasets: dermatology images from HAM10000~\cite{tschandl2018ham10000}, endoscopy images from Kvasir~\cite{pogorelov2017kvasir}, chest X-rays from COVID-QU-EX~\cite{tahir2021covid}, thyroid gland ultrasound from TG3K~\cite{gong2021multi}, and thyroid nodule ultrasound from TN3K~\cite{gong2023thyroid}.
For each dataset, we applied the same target-filtering procedure described in Section~\ref{sec:method} and retained 26, 27, 40, 38, and 40 radiomic targets for HAM10000, Kvasir, COVID-QU-EX, TG3K, and TN3K, respectively, for a total of 171 targets.

\textbf{Evaluation Protocol and Baselines.}
We evaluated each dataset over $50$ random training/calibration/test splits with nominal miscoverage level $\alpha=0.1$ and split proportions $0.4/0.4/0.2$.
We compared four methods: (1) SCP~\cite{lei2018distributionfree}, split conformal prediction using absolute residual scores around the predicted radiomic value; (2) CQR-$\hat{y}$~\cite{romano2019conformalized}, CQR using only the predicted radiomic value $\hat{Y}_i$ as a feature; (3) CQR-Generic~\cite{romano2019conformalized}, CQR using $\hat{Y}_i$, geometry/topology, intensity-distribution, and predicted-radiomics features; and (4) ConRad, which augments CQR-Generic with boundary-uncertainty features.
These baselines add test-time information in stages, allowing us to assess whether boundary uncertainty provides information beyond prediction magnitude and generic image or mask summaries.
For each target, we fit separate linear ridge-regularized lower and upper quantile regressors using the pinball loss, an $\ell_2$ penalty of $5\times10^{-2}$, and Adam optimization for $1200$ iterations with learning rate $5\times10^{-2}$.
We used the same quantile-regression hyperparameters across datasets and targets.

\begin{figure}[t]
    \centering
    \includegraphics[width=\linewidth]{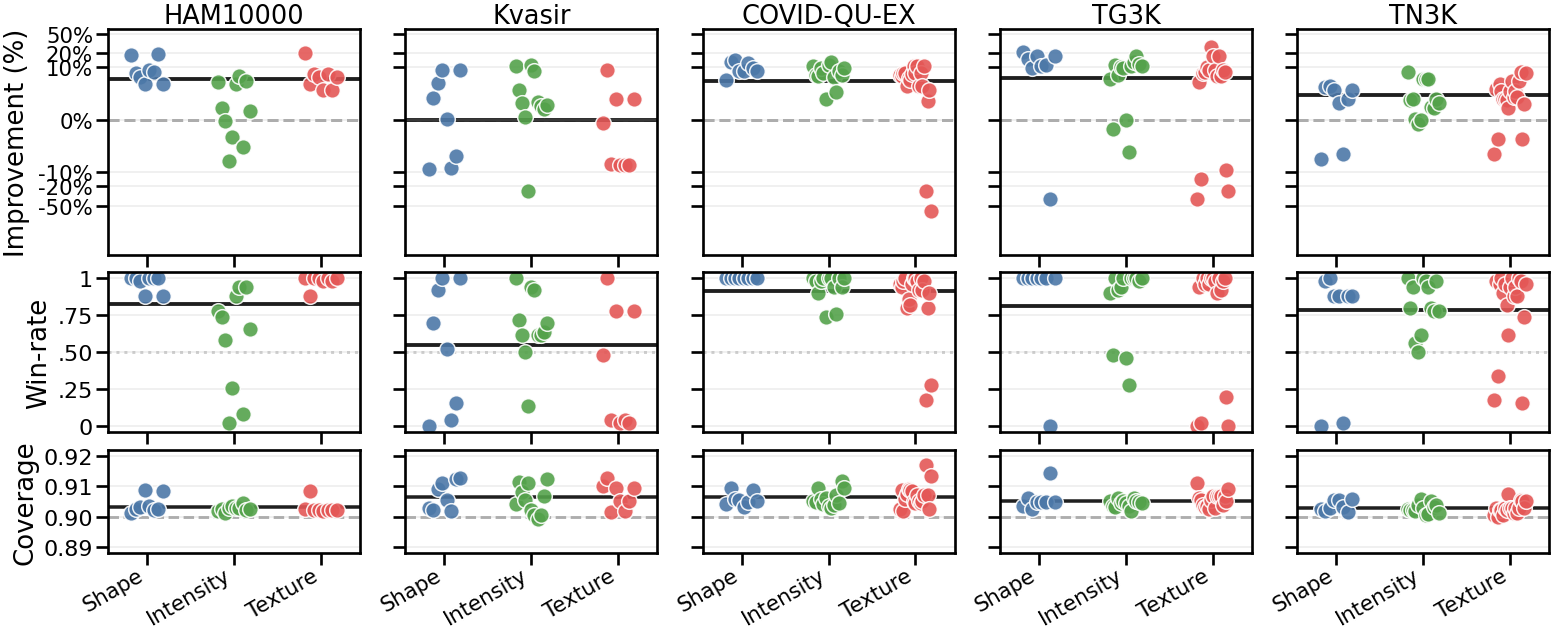}
    \caption{\textbf{ConRad improves feature-level efficiency on most datasets.}
    For $\alpha=0.1$, we show ConRad's mean relative improvement against the strongest baseline (top), win rate against the strongest baseline (middle), and ConRad empirical coverage (bottom) across 50 random splits, stratified by radiomic target family.
    In each subplot, points are radiomic targets, columns are datasets, and colors denote radiomic families.
    Black horizontal lines mark dataset means.}
    \label{fig:featurescatter}
\end{figure}

\textbf{Evaluation Metrics.}
Radiomic targets have different units and numerical scales, so directly averaging raw interval widths across targets is not meaningful.
We therefore evaluate interval efficiency using target-level, scale-normalized comparisons against the strongest baseline.
Let $\mathcal{B}$ denote the set of baseline methods.
For dataset $d$, target $j$, and repeat $r$, define the strongest baseline as
$m^\star_{djr}\in\arg\min_{m\in\mathcal{B}} w_{djr,m}$    
where $w_{djr,m}$ is the mean test interval width for method $m$.
The target-level win rate is
\begin{equation}
    \label{eq:winrate}
   \mathrm{WR}_{dj}
    =
    \frac{1}{R_{dj}}
    \sum_{r=1}^{R_{dj}}
    \mathbf{1}
    \left\{
    w_{djr,\mathrm{ConRad}}
    \le
    w_{djr,m^\star_{djr}}
    \right\},
\end{equation}
where $R_{dj}$ is the number of repeats for target $j$ in dataset $d$, and ties are counted as wins.
To quantify the magnitude of improvement over the best-performing baseline, we report the symmetric relative improvement
\begin{equation}
    \label{eq:relative_improvement}
    \Delta_{dj}
    =
    \frac{200}{R_{dj}}
    \sum_{r=1}^{R_{dj}}
    \frac{
    w_{djr,m^\star_{djr}} - w_{djr,\mathrm{ConRad}}
    }{
    \left|w_{djr,m^\star_{djr}}\right|
    +
    \left|w_{djr,\mathrm{ConRad}}\right|
    } .
\end{equation}
Positive values indicate narrower ConRad intervals; negative values indicate that the best baseline is narrower.
Because interval-width comparisons are meaningful only when coverage is comparable, we also report empirical coverage, $C_{dj,m}=\frac{1}{R_{dj}}\sum_{r=1}^{R_{dj}} C_{djr,m}$ where $C_{djr,m}$ is the empirical test coverage for method $m$ on dataset $d$, target $j$, and repeat $r$.
Coverage values assess whether each method remains close to the nominal marginal coverage level $1-\alpha$.

\section{Results}
\label{sec:results}

Figure~\ref{fig:featurescatter} reports mean relative improvement $\Delta_{dj}$, win rate $\mathrm{WR}_{dj}$, and ConRad empirical coverage $C_{dj,\mathrm{ConRad}}$.
ConRad generally improves target-level interval efficiency, although gains vary across datasets and radiomic families.
Averaged across retained targets, ConRad achieved positive mean relative improvement on HAM10000 (5.47\% [4.93, 5.98]), COVID-QU-EX (4.94\% [4.24, 5.64]), TG3K (5.83\% [5.08, 6.58]), and TN3K (2.03\% [1.86, 2.19]), with corresponding win rates of 82.5\% [79.8, 85.0], 91.3\% [89.5, 93.0], 81.1\% [79.4, 82.7], and 78.6\% [77.0, 80.1].
Kvasir was the weakest dataset, with near-zero mean improvement ($-0.03$\% [$-0.57$, 0.50]) and a lower but still majority win rate (55.3\% [52.6, 58.0]).
Empirical coverage was close to the nominal 90\% level: the minimum mean target-wise ConRad coverage was 89.9\%, the median target-wise coverage ranged from 90.2\% to 90.6\% across datasets, and 170 of 171 targets achieved at least 90\% mean coverage.
The improvements are not explained solely by CQR adaptivity or by generic image-derived features, since the strongest-baseline comparison already includes CQR-Generic.
This suggests that boundary-aware uncertainty provides complementary test-time information for adapting radiomic interval width.

\subsection{Interpretability}

\begin{figure}[t]
    \centering
    \includegraphics[width=\linewidth]{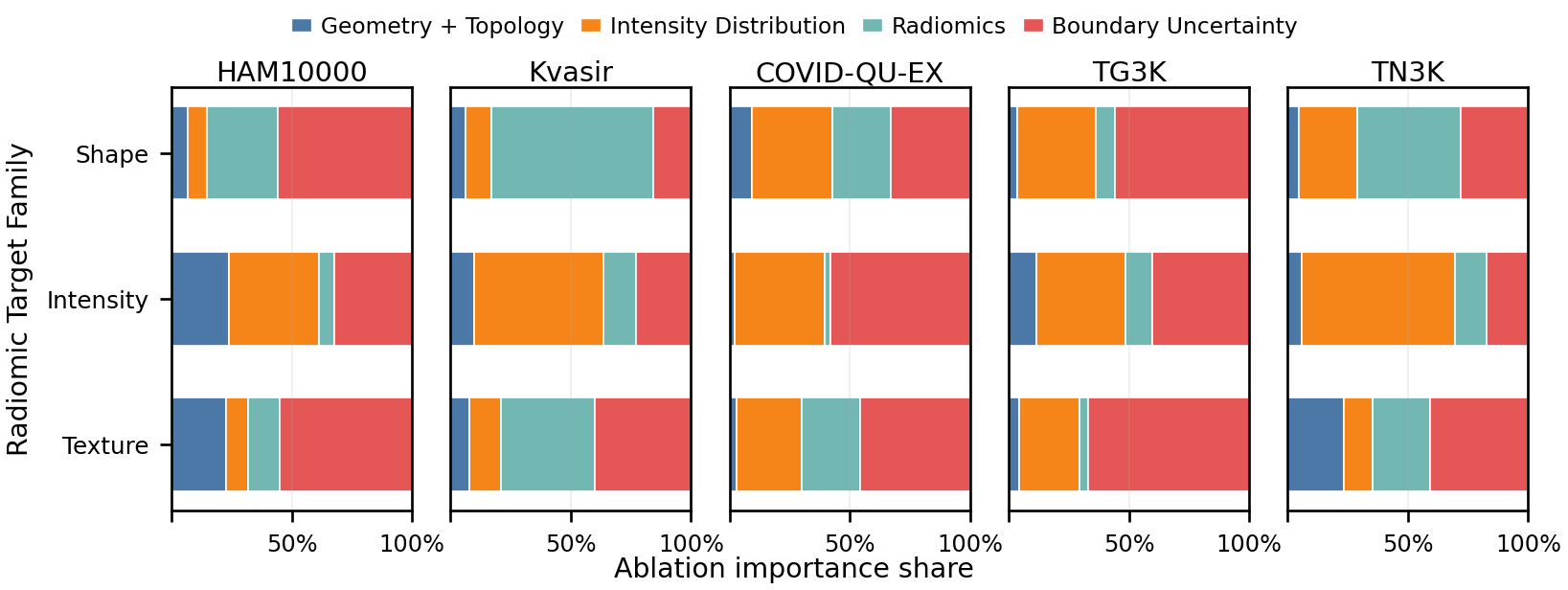}
    \caption{\textbf{Boundary uncertainty features are the largest contributors to ConRad interval efficiency.}
    Each column is a dataset and each row is a radiomic target family.
    Stacked bars show the normalized ablation-importance share from removing one feature family---Geometry + Topology, Intensity Distribution, Radiomics, or Boundary Uncertainty---and refitting and recalibrating the quantile models.}
    \label{fig:interpretability}
\end{figure}

To assess which test-time feature families drive ConRad's efficiency gains, we ran a grouped feature-ablation diagnostic.
For each dataset $d$, radiomic target family $g$, target $j\in\mathcal{J}_{dg}$, and repeat $r\in\mathcal{R}$, we refitted and recalibrated ConRad after removing one feature family $c$.
Let $w_{djr}^{\mathrm{full}}$ be the mean test interval width of the full model and let $w_{djr}^{(-c)}$ be the corresponding width after removing family $c$.
We define positive ablation importance $A_{djrc}$ and normalized importance share $\pi_{dgc}$ as
\begin{align}
    A_{djrc}
    &=
    \max\!\left\{
    200
    \frac{
    w_{djr}^{(-c)} - w_{djr}^{\mathrm{full}}
    }{
    |w_{djr}^{(-c)}| + |w_{djr}^{\mathrm{full}}| + \varepsilon
    },
    0
    \right\},
    \label{eq:ablation_importance}
    \\
    \pi_{dgc}
    &=
    \frac{
    \frac{1}{|\mathcal{J}_{dg}||\mathcal{R}|}
    \sum_{j\in\mathcal{J}_{dg}}\sum_{r\in\mathcal{R}} A_{djrc}
    }{
    \sum_{c'}
    \frac{1}{|\mathcal{J}_{dg}||\mathcal{R}|}
    \sum_{j\in\mathcal{J}_{dg}}\sum_{r\in\mathcal{R}} A_{djrc'}
    } .
    \label{eq:ablation_share}
\end{align}
Here $\varepsilon>0$ is a fixed numerical constant used to avoid division by zero.
If all positive ablation importances are zero for a panel, we set all normalized shares in that panel to zero.
This diagnostic assigns importance only when removing a feature family makes the refitted conformal intervals wider, so it is tied directly to final interval width rather than to model coefficients.

We grouped the test-time features into four families: Geometry + Topology, Intensity Distribution, Radiomics, and Boundary Uncertainty.
In Fig.~\ref{fig:interpretability}, Boundary Uncertainty was the largest contributor on average, accounting for 40.5\% of normalized ablation importance and producing the largest mean positive width increase (4.95\%).
Boundary Uncertainty was the dominant block in 9 of 15 dataset--target-family panels, including all five Texture panels.
Intensity Distribution, Radiomics, and Geometry + Topology accounted for 28.3\%, 21.5\%, and 9.7\% on average, respectively, showing that the value of each feature family is dataset- and target-dependent.
These ablation values are diagnostic and should not be interpreted as universal feature importance across all datasets or radiomic targets.

\section{Conclusion and Discussion}

We introduced \emph{ConRad}, a conformal prediction framework for scalar radiomic targets derived from predicted segmentation masks. ConRad builds target-specific intervals and adapts their width using test-time information from mask geometry, image intensity, predicted radiomics, and segmentation uncertainty. 
Across five 2D datasets and 171 radiomic targets, ConRad improved interval efficiency over baselines while maintaining near-nominal empirical coverage.

This study has limitations. Our experiments focus on 2D datasets, while 3D radiomics may have different stability and acquisition-dependent behavior. ConRad provides feature-wise marginal coverage rather than subgroup-conditional coverage, and its guarantees rely on exchangeability, so dataset shift or scanner shift may require recalibration or additional diagnostics. The method also depends on useful probability maps for boundary-uncertainty features and currently treats each radiomic target separately, leaving multivariate and structured radiomic uncertainty sets as important future directions.

\section*{Acknowledgements}
MC would like to acknowledge support from a fellowship from the Gulf Coast Consortia on the NLM Training Program in Biomedical Informatics and Data Science T15LM007093.
\section*{Code Availability}
Code available at \url{https://github.com/matthewyccheung/conrad}
\bibliographystyle{plainnat}
\bibliography{ref}

\end{document}